\documentclass[showpacs,letterpaper,twocolumn,nofootinbib, nobibnotes,aps,prd]{revtex4}
\pdfoutput=1

\usepackage{graphicx}
\usepackage{amsmath,amsthm,amssymb}

\newcommand{\be}{\begin{equation}}
\newcommand{\ee}{\end{equation}}
\newcommand{\ba}{\begin{eqnarray}}
\newcommand{\ea}{\end{eqnarray}}

\begin{document}

\title{Detection prospects of the Telescope Array hotspot by space observatories}

\author{D.~Semikoz$^{1}$, P.~Tinyakov$^{2}$, M.~Zotov$^{3}$}

\affiliation{$^{1}$APC, Universite Paris Diderot, CNRS/IN2P3, CEA/IRFU, Observatoire de Paris, Sorbonne Paris Cite, 119 75205 Paris, France}
\affiliation{$^{2}$Service de Physique Theorique, Universite Libre de Bruxelles,
CP225, Brussels 1050, Belgium}
\affiliation{$^{3}$D.V. Skobeltsyn Institute of Nuclear Physics, M.V. Lomonosov Moscow State University
(SINP MSU), Moscow 119991, Russia }

\begin{abstract} 
	In the present-day cosmic ray data, the strongest
	indication of anisotropy of the ultrahigh energy cosmic rays
	is the 20-degree hotspot observed by the Telescope Array
	with the statistical significance of 3.4$\sigma$.
	In this work, we study the possibility of detecting such a spot by
	space-based all-sky observatories.  We show that if the detected
	luminosity of the hotspot is attributed to a physical effect and not
	a statistical fluctuation, the KLYPVE and JEM-EUSO experiments would
	need to collect $\sim 300$ events with $E>57$~EeV in order to detect
	the hotspot at the $5\sigma$ confidence level with the 68\%
	probability. We also study the dependence of the detection prospects
	on the hotspot luminosity.  
\end{abstract}

\maketitle


\section{Introduction}
\label{Introduction}


Both cosmic ray protons and nuclei at the highest energies cannot reach us from
cosmological distances due to energy losses on the cosmic microwave background
and infrared backgrounds. The cutoff in the ultrahigh energy cosmic ray (UHECR)
spectrum was predicted by K.~Greisen, G.~Zatsepin, and V.~Kuzmin in
1966~\cite{gzk} and was observed first by the HiRes experiment in
2002~\cite{HiRes} and later confirmed with larger statistical significance by
the Pierre Auger Observatory~\cite{Abraham:2008ru} and Telescope
Array~\cite{AbuZayyad:2012ru}.

The presence of the cutoff in the UHECR spectrum implies that cosmic rays at
the highest energies come from the nearby Universe. At energies
$E\gtrsim60$~EeV one expects that most of the cosmic rays come from local
sources with $z<0.1$. One can hope to find those sources by correlating the
arrival directions of the cosmic ray events with catalogs of astrophysical
sources.

However, charged cosmic rays are deflected from the sky positions of their
sources by both the galactic and intergalactic magnetic fields.  For UHECR
protons with $E\gtrsim60$~EeV, the deflections in the galactic magnetic field
are not large, $\delta_{\rm Gal} \sim 2^\circ (Z/1) (B/\mu G) (60\,\,{\rm
  EeV}/E)$.  According to modern models of the galactic magnetic
field~\cite{Pshirkov:2011um,Jansson:2012pc}, this is true for outside of
the galactic plane in 
most of the sky. Much less clear is the situation with the
extragalactic magnetic fields. Faraday rotation measures of extragalactic
sources set  an upper bound on such fields at a nanoGauss level~\cite{Pshirkov:2015tua}.
Different numerical simulations show contradicting results from very small
deflections $\delta_{\rm extra-Gal} < 1^\circ$ outside of galaxy
clusters~\cite{Dolag:2004kp} to as large as tens of degrees
$\delta_\mathrm{extra-Gal} > 10^\circ$~\cite{Sigl:2004yk}.

Assuming that deflections in the extragalactic magnetic fields are small one
can expect a small-scale (of the order of a few degrees) correlation between arrival
directions of UHECR events and positions of sources located in the large-scale
structure. However, the search for such correlations with point sources was
not successful. First positive hints of correlations with point sources found
in the Auger data~\cite{Abraham:2007bb} were not confirmed by the later data of
both Auger~\cite{PierreAuger:2014yba} and Telescope Array (TA)
experiments~\cite{Abu-Zayyad:2013vza}. At larger angular scales, the results
of the full-sky harmonic analysis~\cite{Aab:2014ila} also suggest that
deflections are larger than what follows from the above
estimate~\cite{Tinyakov:2014fwa}. These negative results indicate either the
presence of a large fraction of intermediate/heavy nuclei at $E\gtrsim60$~EeV
or large extragalactic magnetic fields, or both.

The Auger experiment has detected a change of composition towards heavy nuclei at
high energies~\cite{Abraham:2010yv}. In particular, the most recent measurements
in combination with post-LHC hadronic models show the absence or a small
fraction of both protons and iron at $E>40$~EeV~\cite{Aab:2014aea}. The TA
data are consistent with protons for pre-LHC models, but do not have
sensitivity to distinguish protons from intermediate nuclei at
$E>40$~EeV~\cite{Abbasi:2014sfa}.  On the other hand, joint analysis of both
experiments has shown a consistency of the experimental data on composition
between TA and Auger~\cite{Abbasi:2015czo} within estimated errors. A solution
consistent with currently existing data could be that UHECRs at $E>40$~EeV
are largely composed of intermediate-mass nuclei, and their deflections
prevent us from finding sources by correlating arrival directions with the
source positions at small angles.

Another possibility to look for sources of UHECRs is to use the autocorrelation
function of cosmic rays.  This function is not very sensitive to deflections
in the regular field, which can help to find sources even for nuclei
primaries. The combined data of AGASA and HiRes experiments already indicate a
possible anisotropy at $E>40$~EeV and the 20-degree angular
scale~\cite{Kachelriess:2005uf}. A similar anisotropy was found later in the
Auger data which show an excess in the circle of $18^\circ$ radius centered
near Cen~A~\cite{Abreu:2010ab}. The significance of anisotropy towards Cen~A
has not improved in later data.

Finally, the Telescope Array detected a hotspot in the Northern hemisphere
using the five-year data recorded up to May~4, 2013~\cite{TAhotspot}.  The hotspot
was a cluster of 19 events with energies $>57$~EeV occupying a
$20^\circ$-degree radius circle centered at $\mathrm{R.A}=146.\!^\circ7$,
$\mathrm{Dec}=43.\!^\circ2$, near the Ursa Major cluster of galaxies.  The
\textit{pretrial} statistical significance of the hotspot equals~$5.1\sigma$,
with the \textit{post-trial} probability of it appearing by chance in an
isotropic cosmic ray sky estimated as $3.4\sigma$.  With the additional two
years of data taking, the statistics is not yet enough to confirm the result:
the number of events in the hotspot increased up to~24 but the statistical
significance of the excess remained the same~\cite{TAhotspot-ICRC2015}.

The TA experiment alone can confirm this result in the next few years after
the four-times extension, but an independent confirmation by a different
experiment will be important. In particular, future space-based instruments
like KLYPVE~\cite{Garipov:2015iha,Panasyuk:2015hba} or
JEM-EUSO~\cite{Haungs:2015qfa} can do this job. In this work, we study the
discovery potential of these experiments for an independent detection of the
TA hotspot.


\section{KLYPVE and JEM-EUSO exposure}
\label{exposure}

In order to simulate the distribution of the {\em detected} cosmic ray events
in the arrival directions, one needs to know the exposure of the experiment as
a function of the direction in the sky.
Both KLYPVE and JEM-EUSO are planned for deployment at the
International Space Station. The two instruments are different in design
but employ the same technique for detecting UHECRs.
They will register the near-ultraviolet fluorescent light generated by
secondary particles in extensive air showers born in the
atmosphere by primary UHECRs, and the Cherenkov light reflected at
the surface of the Earth. 
The expected exposure of JEM-EUSO (in nadir observation) was studied
in detail in~\cite{JEM-EUSO:exposure}.
It was shown that the experiment will cover the whole celestial sphere
with the integrated exposure only slightly depending on 
declination~$\delta$ and being uniform with respect to right ascension.
The dependence of exposure on declination obtained
in~\cite{JEM-EUSO:exposure} can be approximately expressed as
\begin{equation}
	R(\delta) = 1 + 0.0185 \sin^4\delta + 0.0192 \sin^6\delta  - 0.006.
\label{eq:Exposure_rel}
\end{equation}
This exposure is nearly uniform over the sphere, with variations not exceeding
a few percent. 
Since both experiments will have the same orbit and share the same
principle of detecting UHECRs, Eq.~(\ref{eq:Exposure_rel}) 
can be used for the KLYPVE mission, too.

Exposure of both detectors depends on the energy of primary particles
but they are expected to be fully efficient at energies above
$\approx50$--$60$~EeV~\cite{Panasyuk:2015hba,K-EUSO:exposure,Olinto-ICRC2015}.
Thus this dependence is not important for what follows since we 
present the results directly in terms of the total number of events with
energies exceeding 57~EeV.


\section{Hypotheses to be tested}
\label{hypotheses}

In this paper, we consider two alternative hypotheses concerning the sky
distribution of UHECRs with $E>57$~EeV: 
\begin{enumerate}
\item[{\bf H0:}] isotropic distribution.
\item[{\bf H1:}] isotropic distribution superimposed with the hotspot 
of a given relative intensity.
\end{enumerate}
Under H0 we generate isotropic events and then modulate their
distribution with the KLYPVE exposure (\ref{eq:Exposure_rel}).%
\footnote{An isotropic flux obeying exposure~(1) can also be
simulated using the standard inverse transformation method.
Our calculations show that both approaches provide identical results
but the first one is more efficient on computer time.}

When generating the events that follow H1 for given hotspot parameters,
we first generate the hotspot events that follow the Gaussian
distribution of a given width and position. Isotropically distributed events
are then added in such a way that the fraction~$f$ of the hotspot events
in the combined set equals the given value.
Finally, the resulting set is modulated with the
exposure~(\ref{eq:Exposure_rel}).

In this paper, we use the hotspot parameters from Ref.~\cite{TAhotspot}. The
right ascension and declination of the center are taken to be $146.\!^\circ7$
and $43.\!^\circ2$ respectively. The uncertainty in the position of the center
is $2.7^\circ$. In Ref.~\cite{TAhotspot}, the hotspot was fitted with the
Gaussian shape plus a uniform background. The width of the spot was found to
be $10.3^\circ$ with the uncertainty of $1.9^\circ$. The amplitudes of the
Gaussian part and the uniform background can be converted into the
fraction~$f$ of the hotspot events as would be seen in the case of a uniform
exposure. This gives $f=0.084$ with the uncertainty $\sigma_f=0.036$.


\section{Prospects of detecting the TA hotspot by space observatories}
\label{probability}


To quantify the discovery potential of the KLYPVE and JEM-EUSO
missions with respect
to the TA hotspot, we calculate how many events should be observed in
order to establish its existence at $5\sigma$ confidence level (C.L.). More
specifically, for a given number of observed events~$N$ we generate many
simulated data samples following~H1. Each sample has the hotspot
parameters picked randomly from a Gaussian distribution centered at the
values measured by the TA~\cite{TAhotspot} with the width equal to the
corresponding standard deviation. The parameters over which the
marginalization is performed include the hotspot position and width. We
do not marginalize over the hotspot intensity; instead, three values are
considered: the central value that corresponds to $f_0=0.084$, and the
optimistic/pessimistic cases $f_\pm=0.084\pm 0.036$.

For each generated sample we calculated the value of the test statistics (TS).
Several test statistics were considered: the number of events~$n_s$ in the
circle of radius $20^\circ$ fixed at the position of the TA hotspot, as well as
the first five spherical harmonic coefficients~$C_l$ with $l=1,\dots,5$. We have
found that the first test statistics is much more sensitive than the others, the
reason being that it incorporates information about the exact hotspot location,
while the harmonic coefficients~$C_l$ are rotationally invariant. In what
follows we present the results for this TS only.

By generating a large number of samples at fixed~$N$ and fixed hotspot
intensity, we constructed a distribution of TS,~$n_s$. From this distribution
we determined the value~$\bar n_s$ of the TS such that 68\% of realizations
have equal or larger value of~$n_s$.

We then generated many samples of~$N$ events corresponding to no-signal
hypothesis H0, calculated the TS for each of them and obtained the
distribution of the TS under~H0. Since we are interested in the $5\sigma$ C.L.,
the number of isotropic samples has to be at least $10^7$. Note, however, that
the distribution of the TS for the isotropic hypothesis is known analytically:
this is just a binomial distribution fully characterized by the ``number of
trials''~$N$ and the ``probability of success in a single trial''~$p_0$. The
latter is just the probability that a single observed event will be found in
the hotspot region. This probability is much easier to calculate numerically;
we have found $p_0 = 0.0302$, including the effect of nonuniform
exposure. Other properties of this distribution, in particular the probability
to have~$n$ or more events in the spot out of~$N$ total, can be calculated
analytically.

\begin{figure}[!ht]
	\centering
	\includegraphics[width=0.45\textwidth,angle=0]{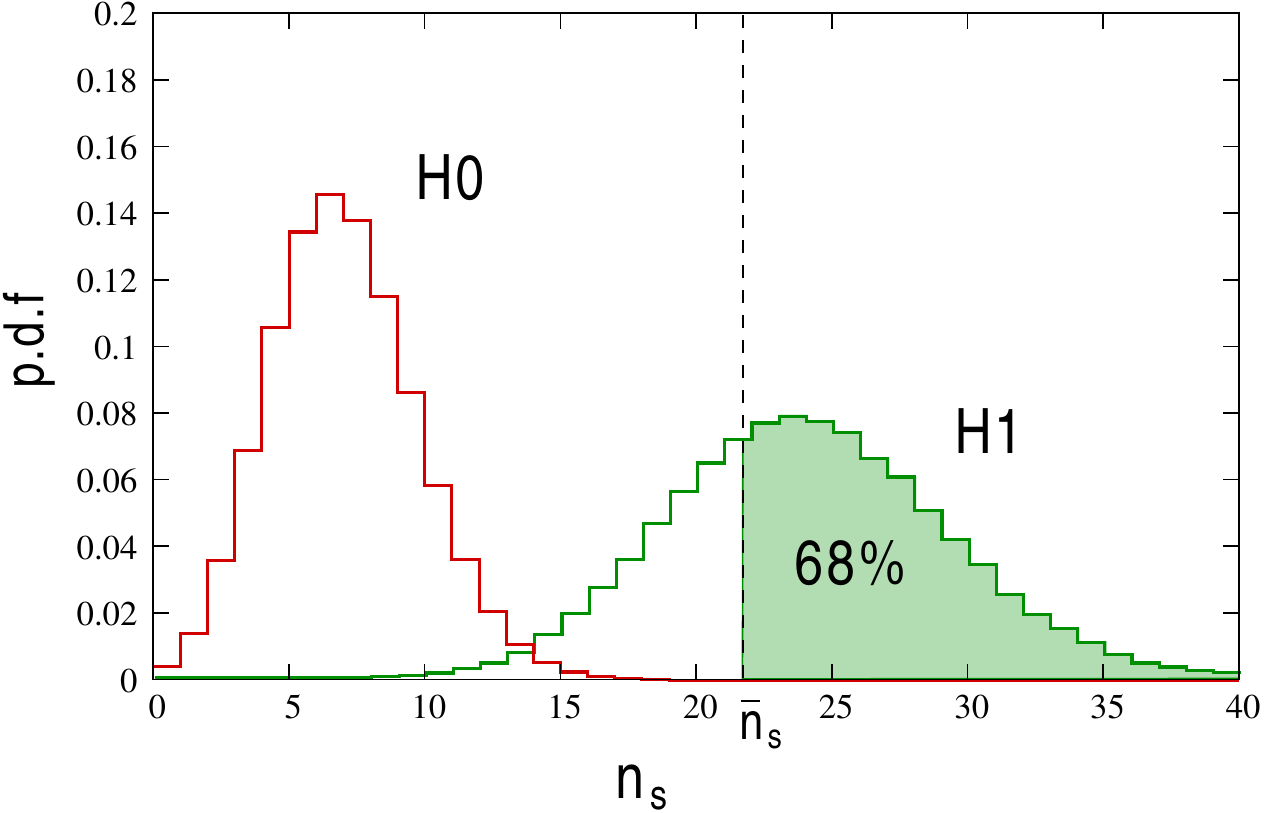}

	\caption{Probability distributions of the number of events~$n_s$ in
          the TA hotspot region for the isotropic distribution (H0) and in
          the case of the hotspot with parameters as determined by TA
          \cite{TAhotspot} (H1). The total number of events is 250. The vertical
          line shows the value~$\bar n_s$ such that 68\% of realizations have
          the signal at least that strong. }
	\label{fig_distr}
\end{figure}

Having obtained $\bar n_s$ for given values of~$N$ and the spot intensity~$f$,
as well as the distribution of the TS under H0, we finally determine the
probability to have, in an isotropic set, the TS~$n_s$ larger than or equal
to~$\bar n_s$ (that is, $\bar n_s$ or more events inside the spot region). This
probability, interpreted as Gaussian and converted into standard deviations,
gives the C.L. at which the isotropy hypothesis H0 can be ruled out in 68\% of
cases for given~$N$ and~$f$. The whole procedure is illustrated in
Fig.~\ref{fig_distr} for particular values of parameters as explained in the
caption.

\begin{figure}[!ht]
	\centering
	\includegraphics[width=0.45\textwidth,angle=0]{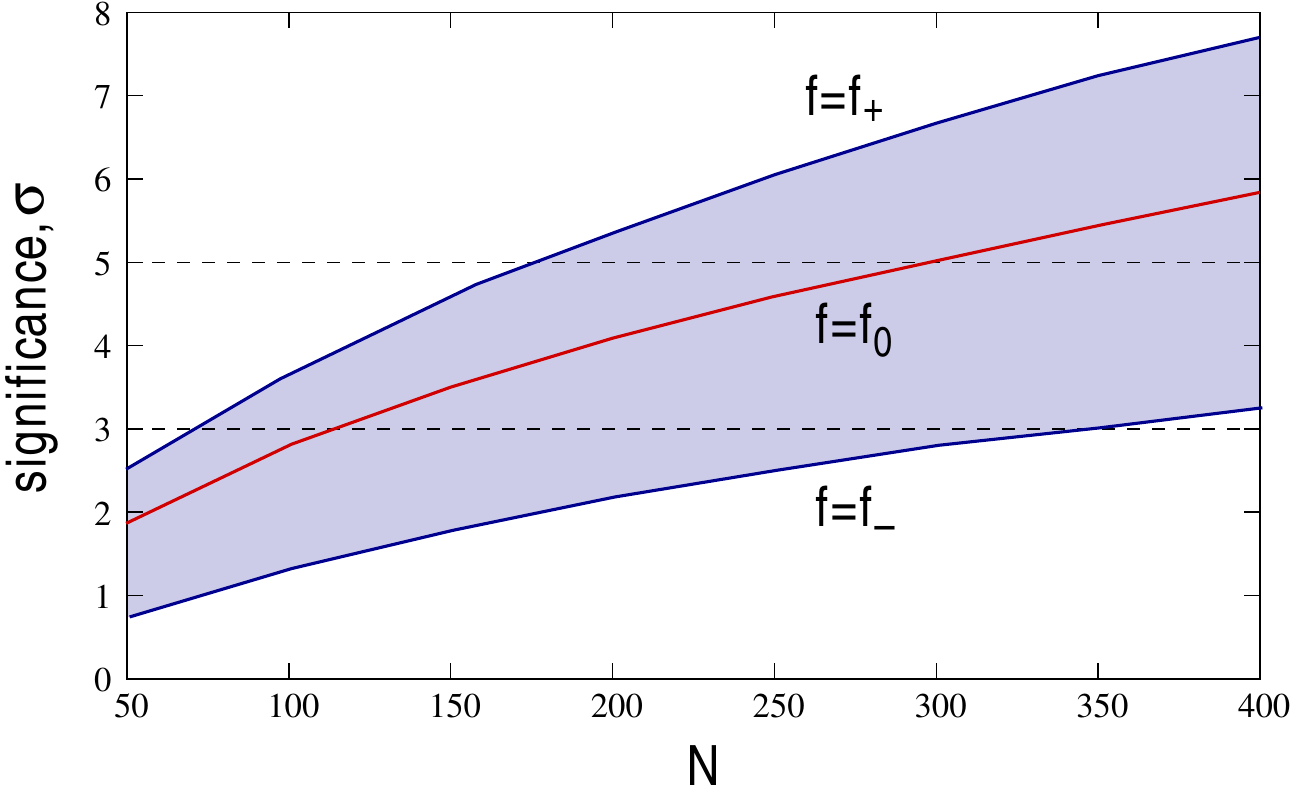}
	\caption{The significance of the isotropy hypothesis rejection
          as a function of the total number of detected events
			 $N$. The central curve (red) $f=f_0$:
			 hotspot brightness as deduced from the five-year
          TA data. The shaded band: corresponding~$1\sigma$ uncertainty.
          Horizontal dashed lines show the $3\sigma$ evidence and
          $5\sigma$ discovery levels.}
	\label{fig_sigma}
\end{figure}

Figure \ref{fig_sigma} shows the dependence of the significance at which
the isotropy hypothesis~H0 can be ruled out as a function of the
observed number of events~$N$ for three values of the spot intensity
$f=f_0,\,f_\pm$ in the best 68\% of cases.

The significance is shown in terms of Gaussian standard deviations~$\sigma$.
Horizontal lines at $3\sigma$ and $5\sigma$ represent the standard evidence and
discovery levels. The red curve in the middle corresponds to the brightness of
the spot as deduced from in the five-year TA data. Upper and lower blue lines
represent the~$1\sigma$  uncertainty of the hotspot brightness. 

If the central value for the hotspot brightness is assumed, then
$3\sigma$ detection can be expected with $\sim120$ events, while a
$5\sigma$ discovery will require the observation of $\sim300$ events
with $E>57$~EeV. In case of the optimistic scenario these numbers
change to~$70$ and $170$, respectively. In case of the pessimistic
scenario the evidence will be obtained with $\sim350$ events, while
the discovery will require accumulation of $\sim1000$ events with
$E>57$ EeV.

Will KLYPVE or JEM-EUSO be able to register the necessary number of events?  It
was estimated recently that with the annual exposure $\sim5\times10^4$~km$^2$~sr
above $\sim60$~EeV, JEM-EUSO will collect 429 events/yr, or about 2,145 events
in five years~\cite{Olinto-ICRC2015}.  In a similar fashion, one can estimate
that with the annual exposure $\sim1.2\times10^4$~km$^2$~sr, KLYPVE will detect
more than 100 events every year of operation, and more than 600 events during
its planned lifetime.  Thus, both experiments have a strong discovery potential
to detect the TA hotspot.

\section{Conclusions}
\label{conclusions}

In this work, we studied the possibility of the TA hotspot detection
by future space experiments like KLYPVE and JEM-EUSO. We have seen
that the perspectives of the hotspot detection depend strongly on the
actual signal strength. If the mean strength derived from the five-year TA
data is assumed, with $\sim300$ observed events with $E>57$~EeV the
space observatories will have a 68\% chance of the $5\sigma$
discovery.
The number of events required for that would be $\sim1000$
in the case of the pessimistic scenario.

With its huge annual exposure (almost an order of magnitude larger than 
that of the Pierre Auger Observatory) and the planned five-year operation
time, JEM-EUSO has excellent opportunities for confirming the existence
of the TA hotspot at high confidence level.
In six years of operation, KLYPVE will have the total exposure
approximately~$1/3$ of JEM-EUSO, and thus it also has a strong
discovery potential, especially in the case in which the five-year flux registered
by the Telescope Array persists.

\section*{Acknowledgements}
The work was done with  partial financial support from
the Russian Foundation for Basic Research Grant No.\ 13-02-12175-ofi-m.

\bigskip

Note added: Recently, we became aware
of a similar work reported at the 18th JEM-EUSO International
Meeting (Stockholm, December 7--11, 2015)~\cite{Shinozaki}.
As far as we understand, the results presented there
were obtained in a different fashion but are close to our own.


\end{document}